\documentclass{article}
\usepackage[utf8]{inputenc}
\usepackage[colorlinks = true,
            linkcolor = blue,
            urlcolor  = blue,
            citecolor = blue,
            anchorcolor = blue]{hyperref}

\def\beq{\begin{equation}}
\def\eeq#1{\label{#1}\end{equation}}
\def\eeqn{\end{equation}}

\newenvironment{Eqnarray}%
   {\arraycolsep 0.14em\begin{eqnarray}}{\end{eqnarray}}
\def\beqa{\begin{Eqnarray}}
\def\eeqa#1{\label{#1}\end{Eqnarray}}
\def\eeqan{\end{Eqnarray}}

\let\bar=\overbar

\def\lsim{\mathrel{\raise.3ex\hbox{$<$\kern-.75em\lower1ex\hbox{$\sim$}}}}
\def\gsim{\mathrel{\raise.3ex\hbox{$>$\kern-.75em\lower1ex\hbox{$\sim$}}}}

\def\del{\partial}
\def\Dslash{\not{\hbox{\kern-4pt $D$}}}
\def\dslash{\not{\hbox{\kern-2pt $\del$}}}
\def\pslash{\not{\hbox{\kern-2pt $p$}}}
\def\ETmiss{\not{\hbox{\kern-4pt $E$}}_T}

\def\Dlr{\mathrel{\raise1.5ex\hbox{$\leftrightarrow$\kern-1em\lower1.5ex\hbox{$D$}}}}

\def\MSB{{\bar{M \kern -2pt S}}}
\def\msb{{\bar{\scriptsize M \kern -1pt S}}}

\def\drb{{\bar{\scriptsize D \kern -1pt R}}}

\def\Title#1{\begin{center} {\LARGE #1 } \end{center}}
\def\Author#1{\begin{center}{ \sc #1} \end{center}}
\def\Address#1{\begin{center}{ \it #1} \end{center}}
\def\andauth{\begin{center}{and} \end{center}}

\newenvironment{Abstract}{\begin{quotation} \begin{center}
                       ABSTRACT
     \end{center}\bigskip  }{\end{quotation}}

\newcommand\snowmass{\begin{center}\rule[-0.2in]{\hsize}{0.01in}\\\rule{\hsize}{0.01in}\\
\vskip 0.1in Submitted to the  Proceedings of the US Community Study\\ 
on the Future of Particle Physics (Snowmass 2021)\\ 
\rule{\hsize}{0.01in}\\\rule[+0.2in]{\hsize}{0.01in} \end{center}}

\begin{document}
\snowmass
\Title{Towards an HPC Complementary Computing Facility}
\Author{Kenneth Herner} 
\Author{Michael Kirby}
\andauth
\Author{Steven Timm}
\Address{Fermi National Accelerator Laboratory, Batavia, IL, 60510}
\date{March 2022}

\begin{Abstract}
This Letter considers the design for computing facilities that are complementary to the leadership class High Performance Computing (HPC) facilities. This design envisions a future where funding agencies are allocating greater resources for leadership class facilities and these facilities will provide a significant part of the total compute cycles for HEP Experiments. While a leadership class facility (LCF) may provide cycles and advanced architectures, the facility does not necessarily provide all of the services needed to help HEP users make the best use of the HPC facility, as well as the services needed to provide computing for workflows that are not a good fit for the HPC facilities. This Letter outlines some of the necessary components of a facility designed to provide those services and capabilities.
\end{Abstract}

\section{Introduction}
As leadership class facilities become a larger and larger part of the distributed computing model of HEP experiments, it is important for the current and future generations of neutrino experiments to take advantage of these facilities as their workflows have a number of computing tasks which have been shown to be good use cases for GPU- and accelerator-based computing. 
These workflows include, but are not limited to, processing data from Liquid Argon Time Projection Chambers, pattern recognition, particle identification, and final model parameter fits. 
Early studies have shown good results from a mixed CPU-GPU analysis~\cite{sonic}, which is a natural fit for the next generation of DOE leadership-class machines. 
With the HEPCloud project at Fermilab, the Fermilab Scientific Computing Division has been working on behalf of experiments to provision resources at LCFs and other supercomputing centers since 2017~\cite{hepcloud,hepcloud2}. 

At the same time, considerable development is necessary in order to transition from current HTC computing models and to efficiently access resources at LCFs. 
Most workflows directed to LCFs have required limited I/O and external connectivity, (e.g. Monte Carlo simulation) and produced a modest amount of output.
Additionally some small or late-in-life experiments will lack sufficient effort to modify their computing model to efficiently utilize HPC resources.
To enable more workflows to access HPC computing facilities, particularly those LCFs which have very restricted network input and output, a complementary facility should exist to provide a series of needed capabilities.
All of these capabilities currently exist in some form but need to be improved in terms of scalability and accessibility. 

\section{Characteristics of a complementary facility}
We envision that a complementary facility could serve as a central nexus to drive development efforts on the following components and capabilities to help users effectively utilize HPC facilities:
\begin{itemize}
\item {\bf Code distribution:} The CVMFS system \cite{cvmfs} is currently used at most grid computing sites. However the complete lack of outside network connectivity on worker nodes at some LCFs make this an untenable solution. Containerized execution environments might solve this problem, though facilities have very restrictive container policies making standard tools inoperable. There are solutions that may potentially address both issues (e.g. cvmfsexec in user space), but there is still a need for more integration and scale testing in the field. A complementary facility could house such development efforts and drive common solutions across multiple experiments.
\item {\bf Databases:} One of the most challenging aspects of distributed computing at LCFs is access to databases and distribution of reconstruction configuration constants. There is a need to understand both access patterns (especially on HPC platforms with little to no external connectivity on worker nodes) and whether or not the FronTier model \cite{frontier,frontier2} of distributed databases will continue to be applicable within the architecture of LCFs. 
\item {\bf Data movement:} Rucio \cite{rucio} provides the tools needed to move the data from storage elements to LCFs on demand, but development is still needed for the interfaces from the batch systems and workflow management systems to invoke the tools automatically to transfer data to and from the facilities. The data management tools must also adapt to LCF storage systems of varying I/O and network bandwidth and sync the data in a timely way. There is potential to explore the possibility of a data service that analysis frameworks can call, which would make the data source and output locations transparent to the end user job and independent of network topology. Finally, exploration into the meaning of data movement in an environment where some or all of the data may be stored in non-file-like object stores is critical to match with the storage elements at LCFs. There is also a need for a host lab storage system with I/O and network bandwidth sufficient to send and receive the both production data and user outputs from the LCF, especially for such tasks as interactive analysis which is often difficult to predict or schedule in advance.
\item {\bf Monitoring:} Stakeholders want to know accurate information of the uptime and year to date usage of the HPC facilities. They also need good metrics of their job efficiency on these facilities. Each HPC facility has a different monitoring system to which the API frequently changes. There is thus a need to develop a unified monitoring system capable of presenting monitoring data from a variety of LCFs in a common format, relieving the end user of the responsibility of mastering separate systems at each facility.
\item {\bf Campaign- and Data-aware Workload Management:}  In the last decade HEP experiments have been using a late-binding model where jobs are matched on a per-job basis to remote sites. We would like to add the capacity to define a “campaign” of related jobs and then to schedule them all at a single remote site, transferring the execution tracking of those jobs to the remote site while the campaign is running. Such a system could reserve a large part of the facility and dedicate some of the nodes to be data servers of a memory-based object store to the rest of the nodes. Additionally, many batch jobs at remote sites require a specific version of “pileup” files and/or “overlay” files to be present at the site to avoid the inefficiency of reading or streaming them through bandwidth-limited network connections to individual worker nodes; this system could appropriately steer such jobs to sites that host such data, reducing the overall campaign data movement requirements. Again, as much commonality across experiments as possible is a highly desired feature in such tools as it reduces overhead when moving between experiments.
\item {\bf Test and compilation hardware:} Developers need local access to machines having architectures similar to the HPC systems so that they can find the appropriate way to build code to best utilize the LCF architecture. Since LCF machines are often highly specialized, a one-size-fits-all approach will likely not work here. The suite of test and compilation hardware would need to encompass a wide variety of architectures and computing paradigms (i.e. CPU-dominated, GPU-dominated, etc.)
\end{itemize}

\section{Conclusion}
With the increased availability and importance of LCFs in computing models, there is a significant need for the development and improvement of HPC complimentary computing facilities.
These facilities are essential to the effective utilization of LCFs and bridge the gap between experimental workflows' resource needs and the limitations of LCF infrastructure. 
The development and deployment of these facilities are especially important for experiments with limited effort for computing development, and commonality, both in interfaces to LCFs and in tool functionality across experiments, will be a key feature of such facilities that will help to maximize science output.

\end{document}